\documentstyle[12pt]{article}
\newcommand{\be}{\begin{equation}}
\newcommand{\ee}{\end{equation}}
\newcommand{\bea}{\begin{eqnarray}}
\newcommand{\eea}{\end{eqnarray}}
\newcommand{\beas}{\begin{eqnarray*}}
\newcommand{\eeas}{\end{eqnarray*}}
\newcommand{\comp}{\mbox{\scriptsize $\circ$}}

\newcommand{\<}{\langle}
\renewcommand{\>}{\rangle}
\renewcommand{\l}{\left}
\newcommand{\r}{\right}

\newcommand{\id}{{\bf 1}}

\newtheorem{theorem}{Theorem}[section]

\newtheorem{corollary}{Corollary}[theorem]

\newtheorem{nearremark}{{\it Remark}}[section]
\newenvironment{remark}{\begin{nearremark} \rm}{\end{nearremark}}
\newcounter{stokes}

\newenvironment{proof}{\vspace*{2ex}\noindent {\em Proof:} }{$\Box$ \\[2ex]}

\renewcommand{\dim}{\mbox{dim }}
\newcommand{\kerpr}{\ker P_{\! r}}
\begin{document}
\newcommand{\dista}{2ex}
\newcommand{\distb}{7ex}
\newcommand{\distc}{.25ex}
\newcommand{\distd}{2in}
{ \raggedleft \flushright \hfill To appear in {\em J. Math. Phys.} Spring
1996}   							\\[1in]
\begin{centering}{\large\bf Topological Sectors and Measures on Moduli
Space in Quantum Yang-Mills on a Riemann Surface} 		\\[\distd]
Dana Stanley Fine 			\\ 
Department of Mathematics 		\\ 
University of Massachusetts Dartmouth	\\ 
North Dartmouth, MA 02747 		\\ 
dfine@umassd.edu			\\[.5in]

\begin{quote}\small
Previous path integral treatments of Yang-Mills on a Riemann surface
automatically sum over principal fiber bundles of all possible
topological types in computing quantum expectations. This paper
extends the path integral formulation to treat separately each
topological sector. The formulation is sufficiently explicit to
calculate Wilson line expectations exactly.  Further, it suggests two
new measures on the moduli space of flat connections, one of which
proves to agree with the small-volume limit of the Yang-Mills measure.
\copyright {\em 1996 American Institute of Physics.}
\end{quote}
\end{centering}

\noindent {\bf Index:} 02.90.+p, 02.40.Vh, 11.15.Tk, 11.10.Kk
\newpage

\renewcommand{\thesection}{\Roman{section}}\vspace{-4ex}
\noindent {\bf Running title:} Quantum Yang-Mills on a Riemann Surface
\section{Introduction}
\renewcommand{\thesection}{\arabic{section}} 
In Refs.~\cite{me1}
and~\cite{me2}, we use the path integral formalism to evaluate quantum
expectations of Wilson lines in the Yang-Mills theory on $G=SU(N)$
product bundles over Riemann surfaces of any genus. Other approaches
to these expectations include Sengupta's stochastic quantization of
Ref.~\cite{sengupta2} and Blau \& Thompson's use of the Nikolai map to
simplify the gauge-fixed path integral in Ref.~\cite{blauthomp}.
Witten, in Ref.~\cite{witten1}, derives these expectations
combinatorially and via a Hilbert-space approach using axioms of
quantum field theory for any simple Lie group $G$. He notes that these
results, by contrast with an approach based on Verlinde's formula,
automatically sum over all topological types. In the more recent
treatment of Ref.~\cite{witten2}, he describes how to modify his
Hilbert space approach to treat separately each topological sector.
Likewise, in Refs.~\cite{senguptapreprint} and~\cite{senguptanato},
Sengupta extends the stochastic quantization to non-simply connected
$G$ in a manner which treats separately each topological sector.

In this paper, we analyze the path integral formulation of Yang-Mills
on a principal $G$-bundle of fixed topological type. While doing so,
we augment Witten's results on the partition function by including the
insertion of Wilson lines. These agree with Sengupta's recent results.
Our main focus, however, is on the path integral itself. There is a
recognized scarcity of field theories wherein the path integral can be
evaluated (or interpreted) non-perturbatively to obtain something
approaching a well-defined measure on configuration space. With the
present extension of the detailed account of how to perform such an
evaluation (in a manifestly gauge-invariant fashion), we hope to
contribute to the eventual rigorous understanding of gauge-theoretic
path integration.

As an immediate benefit, in addition to the above-mentioned extension
of Witten's results, we obtain new measures on the moduli space of
flat connections. For genus $g \geq 1$, the moduli space $\cal M$ of
(irreducible) flat connections is a finite-dimensional manifold.
Witten uses the small-volume limit of Yang-Mills to compute the volume
of $\cal M$, which he shows agrees with that defined by the symplectic
volume form on $\cal M$.  In Ref.~\cite{forman}, Forman extends this
agreement to one between measures on $\cal M$. Our approach suggests
two new measures on $\cal M$. We show that one of these, which was
arrived at independently by King and Sengupta in Ref.~\cite{ks}, is
equivalent to the small-volume limit of the Yang-Mills measure.

This paper is organized as follows:             \\
\begin{description}{\addtolength{\labelwidth}{10em}}
\item[Section~\ref{review}]{gives a brief summary of results for $G=SU(N)$
product bundles.}
\item[Section~\ref{top}]{describes the modifications required to keep track
of individual topological sectors in treating bundles with non-trivial
topology.}
\item[Section~\ref{measures}]{introduces the new measures on $\cal M$ and
compares them with the small-volume limit of the Yang-Mills measure.}
\end{description}

\renewcommand{\thesection}{\Roman{section}}
\section{Review of the product case}            
\renewcommand{\thesection}{\arabic{section}}
\label{review}
Let $P$ be a principal bundle with symmetry group $G$ over a Riemann
surface $M$ with a base point $m \in M$. Let ${\cal A}$ denote the
space of connections on $P$, and let ${\cal G}_{m}$ denote the group
of gauge transformations which are the identity on the fiber over $m$.
In Ref.~\cite{me2}, we
describe ${\cal A / G}_{m}$ as itself a principal bundle with
affine-linear fiber over $Path^{2g}\,G$, the space of $2g$-tuples of
paths in $G$ subject to the following relation on the $4g$ endpoint
values
$\l\{\alpha_i(0),\alpha_i(1),\beta_{i}(0),\beta_i(1)\r\}_{i=1}^{g}$:
\be
\prod_{i=1}^{g}\alpha_i(0)\beta_i(1)^{-1} \alpha_i(1) ^{-1} \beta_i(0)
= \id.                          \label{oldrelation}
\ee
Here, successive factors multiply from the right, and $g \geq 1$. The
projection $\xi : {\cal A / G}_{m} \rightarrow Path^{2g}\,G$ is obtained
from holonomy about a certain one-parameter family of closed paths in $M$
determined by a choice of fundamental domain and $2g$ generators $\l\{a_i,
b_i\r\}_{i=1}^{g}$ of $\pi_1(M)$; the unusual naming of the components of
the elements of $Path^{2g}\,G$ reflects this genesis. In the genus-$0$ case,
the base space is simply $\Omega G$, the space of based loops in $G$.

As the holonomies giving rise to $\xi$ enter into the account of
topological sectors, we review them in more detail here. Let $D$ be a
fundamental domain for $M$ and consider a family of paths from the
base point $m$ to the boundary $\partial \! D$ of $D$ such that every
point of $M$ except $m$ lies on exactly one such
path. Call these radial paths. Let $p$ be a point of an edge of
$\partial \! D $ corresponding to a generator of the fundamental group
of $M$, and let $p ^{-1} \in \partial \! D$ denote the point
corresponding to $p$ in the identification of the edges of $\partial
\! D$. Consider, for a given connection $A$, the holonomy of the closed path
originating at $m$, following the radial path to $p$ and returning from the
radial path through $p ^{-1}$. See figure 1.  Relative to a fixed point on
the fiber over $m$, this holonomy determines an element of $G$ which we
denote by $\alpha_i(p)$ if $p$ is in the edge corresponding to the generator
$a_i$ ($\beta_i(p)$ if the corresponding generator is $b_i$). As $p$ varies
along its edge, $\alpha_i(p)$ describes a path in $G$. The $2g$-tuple of
such paths in $G$ described as $p$ moves among the edges corresponding to
generators is $\xi\l([A]\r) \in Path^{2g}\,G$. The endpoint relation
reflects the fact that the radial paths to the vertices of $\partial \! D$
are each traversed twice. In particular, the concatenation of paths whose
holonomy is $\prod_{i=1}^{g}\alpha_i(0)\beta_i(1)^{-1} \alpha_i(1) ^{-1}
\beta_i(0)$ is a contractible path containing no area, so the holonomy
is the identity in accordance with Eq.~\ref{oldrelation}. The fiber of $\xi$
is the affine-linear space $\kerpr$ of Lie-algebra-valued one-forms which
vanish in the radial direction.

In this picture, the path integral for the expectation of some
function on ${\cal A / G}_{m}$, in the Yang-Mills measure $\mu$, is an
iterated integral over the linear fibers and the base $Path^{2g}\,G$.
The integral over the fibers is Gaussian. Performing this integral
yields a path integral expression for the push-down measure
$\xi_*(\mu)$. The main result of Ref.~\cite{me2} is that
$\xi_*(\mu)$ is the product of Wiener measures on the components
$\{\alpha_i,\beta_i\}$ of the elements of $Path^{2g}\,G$, conditioned
to satisfy the endpoint relation of Eq.~\ref{oldrelation}.  This
measure is computed by integrating products of heat kernels on $G$.
For example, when $g=1$, the partition function $Z$ is given by
\[
Z = \int H\l(\alpha(0)^{-1} \alpha(1);2\rho_a\r)H\l(\beta(0)^{-1}
\beta(1);2(\rho-\rho_a)\r),
\]
where the integral is over all possible values of $\alpha(0)$,
$\alpha(1)$ and $\beta(0)$, and $\beta(1)$ is determined by the relation
$\alpha(0)\beta(1)^{-1} \alpha(1) ^{-1} \beta(0) = \id$. Here, $H$ is
the heat kernel, $\rho$
is the total area of the surface, and $\rho_a$ is the area bounded by
the pair of paths whose holonomies determine $\alpha(0)$ and
$\alpha(1)$. The convolution property of the heat kernel reduces this
expression to
\[
Z = \int H\l((\beta(0) ^{-1} \alpha(1) ^{-1} \beta(0)
\alpha(1);2\rho\r) \, d \beta(0) d \alpha(1).
\]
More generally, for genus $g \geq 1$, 
\be                                     \label{fixedleft}
Z = \int H\l(\prod_{i=1}^{g} x_i y_i x_i ^{-1} y_i ^{-1};2\rho\r) dx
dy.
\ee
Since 
\be                                     \label{heatkernel}
H(x;t) =
\sum_{\mu}(\dim \mu)\chi_{\mu}(x)e^{-2c(\mu)},
\ee
where $\chi_{\mu}$ is the character of the representation
$\mu$, and $c$ denotes the
quadratic Casimir, the partition function is  the sum
\[
Z = \sum_{\mu} \frac{e^{-4c_{\mu}\rho}}{\l(\mbox{dim }\mu\r)^{2g -
2}}.
\]
All sums are over the irreducible representations of $G$.  The induction
step in reducing the integral expression of Eq.~\ref{fixedleft} to the above
sum is given by a pair of integral (orthogonality) relations among the
characters:
\beas
\int \chi_{\mu}\l(wxyx ^{-1} y ^{-1}\r) \, dy &=&
\frac{\chi_{\mu}\l(wx\r)\chi_{\mu}\l(x
^{-1}\r)}{\dim \mu},            \mbox{\hfill and}       \\      
\int \chi_{\mu}\l(wx\r)\chi_{\mu}\l(x
^{-1}\r) \, dx &=& \frac{\chi_{\mu}(w)}{\dim \mu}.
\eeas

This formalism also treats the insertion of Wilson lines. For example, the
expectation of an unknotted Wilson line, in the representation $\mu$, given
by the trace of holonomy about a non-contractible, homotopically non-trivial
loop $C$, is
\[
\l< {\cal W}_{\mu} \r> = \frac{1}{Z}\int \chi_{\mu}(x_1)H\l(\prod_{i=1}^{g}
x_i y_i x_i ^{-1} y_i ^{-1}; 2\rho\r) \, dxdy. 
\]
As a sum over characters, this is 
\[
\l< {\cal W}_{\mu} \r> = \frac{1}{Z} \sum_{\nu}
D_{\mu\nu\nu}\frac{e^{-4c_{\nu}\rho}}{(\dim \nu)^{2g - 2}},
\]
where $D_{\mu\nu\nu}$ is the Clebsch-Gordan coefficient:
$\chi_\mu(x)\chi_\nu(x) = \sum D_{\mu\nu\sigma}\chi_\sigma (x)$.

To make sense of the path integral, we restricted connections in
${\cal A}$ to  have finite Yang-Mills action and to satisfy a
continuity restriction. Without
a refinement of this restriction, bundles $P$ of different topological
types are indistinguishable. Thus, although nominally we treated only
product bundles, our results, when naively extended to
non-simply-connected symmetry groups, correspond to a sum over all topological
types.

\renewcommand{\thesection}{\Roman{section}}
\section{Separating the topological sectors}            \label{top}
\renewcommand{\thesection}{\arabic{section}}
To sort out the topological sectors in the case where $G$ is not
simply-connected and $g \geq 1$, let $\widetilde{G}$ be the covering group
of $G$. Then $G = \widetilde{G}/\Gamma$, where $\Gamma$
is a subgroup $\l\{\id, u_1, \cdots,u_n\r\}$  of the finite center of
the simply-connected Lie group $\widetilde{G}$. The topological type
of $P$ can be characterized as follows: Consider holonomy by a flat
connection about  contractible paths in $M$. As  elements of $G$,
these must be the identity. However, if we lift to $\widetilde{G}$,
these holonomies, though equal to each other, can be any element $u$ of
$\Gamma$. This element defines the topological type of $P$. 

The description in Sec.~\ref{review} of $\xi$ goes through as
before to yield the same endpoint relation for $Path^{2g}\,G$.
However, if we attempt to lift from $G$ to $\widetilde{G}$, the
holonomy of the right-hand side of the relation in Eq.~\ref{oldrelation} will
be replaced by
the element $u$ of $\Gamma$ labelling the topological type of $P$. This
follows from the fact that the concatenation contains no area (so the
holonomy is the same as for a flat connection) and is contractible.
We have thus proven the required refinements of the main theorems
of Refs.~\cite{me1} and~\cite{me2}:
\begin{theorem} On a principal fiber bundle of topological type $u$,
${\cal A / G}_{m}$ is itself a fiber bundle with projection $\xi$ and
affine-linear fiber. The base space $Path^{2g}\,\widetilde{G}$
consists of all $2g$-tuples of paths in $\widetilde{G}$, subject to the
relation 
\be                             \label{newrelation}
\prod_{i=1}^{g}\alpha_i(0)\beta_i(1)^{-1} \alpha_i(1) ^{-1} \beta_i(0)
= u.                    
\ee
\end{theorem}

\begin{theorem}
The push-down measure $\xi_{*}(\mu)$ is the product of Wiener measures
on the components of each element of $Path^{2g}\,\widetilde{G}$,
conditioned to satisfy Eq.~\ref{newrelation}. 
\end{theorem}

These allow us to calculate the partition function and the expectation
of Wilson lines on a bundle of type $u$ over a surface of genus $g$.
For example, the calculation of
the partition
function for a bundle of type $u$ on the torus ($g = 1$) begins as
before:
\[
Z(u) = \int H\l(\alpha(0)^{-1} \alpha(1);2\rho_a\r)H\l(\beta(0)^{-1}
\beta(1);2(\rho-\rho_a)\r).
\]
Now, however, $\beta(1)$ is determined by the relation
$\alpha(0)\beta(1)^{-1} \alpha(1) ^{-1} \beta(0) = u$. Thus,
\[
Z(u) = \int H\l(\beta(0) ^{-1} \alpha(1) ^{-1} \beta(0)
\alpha(1) u ^{-1} ;2\rho\r) \, d \beta(0) d \alpha(1).
\]
More generally, for genus $g \geq 1$, 
\[
Z(u) = \int H\l(\prod_{i=1}^{g} x_i y_i x_i ^{-1} y_i ^{-1} u ^{-1};2\rho\r)
dx dy. 
\]
In all these expressions the integrals are over copies of $\widetilde{G}$ and
$H$ is the heat kernel on $\widetilde{G}$.
Decomposing $H$ as a sum of characters (of representations of
$\widetilde{G}$) according to Eq.~\ref{heatkernel},
\be                                                     \label{partitionfcn}
Z(u) = \sum_{\mu} \frac{e^{-4c_{\mu}\rho}}{\l(\mbox{dim }\mu\r)^{2g -
2}}\, \frac{\chi_{\mu}(u ^{-1})}{\l(\mbox{dim }\mu\r)}.
\ee
This agrees with Witten's results from Ref.~\cite{witten2}[Sec. 4] except for
a constant factor depending on $G$ and $g$ but not on $\rho$.

Incorporating Wilson lines given by parallel transport about the
radial paths used to define $\xi$ is as straight-forward as in the
product case. For instance, the expectation of the Wilson line 
$\chi_{\sigma}(\alpha_i(p))$, given by the trace of holonomy about a
non-contractible, homologically non-trivial loop, is
\beas
\l< \chi_{\sigma}(\alpha_i(p)) \r>& = & \frac{1}{Z(u)} \int
\chi_{\sigma}(x)H\l(\prod_{i=1}^{g} x_i y_i x_i ^{-1} y_i ^{-1} u
^{-1};2\rho\r) dx  dy                              \\
&=&\sum_{\mu}D_{\sigma\mu\mu}\frac{e^{-4c_{\mu}\rho}}{\l(\mbox{dim
}\mu\r)^{2g - 2}}\, \frac{\chi_{\mu}(u ^{-1})}{\l(\mbox{dim
}\mu\r)}.
\eeas
By constrast, the expectation of a Wilson line coming from a
contractible loop is
\beas
\l< \chi_{\sigma}\l(\alpha_i(p_1)^{-1} \alpha_i(p_2)\r) \r>& = &
\frac{1}{Z(u)} \int
\chi_{\sigma}(\bar{x})H(\bar{x};2\rho_a)                        \\
&& \times H\l(\bar{x}^{-1} \prod_{i=1}^{g} x_i y_i x_i ^{-1} y_i ^{-1}
u ^{-1};2(\rho-\rho_a)\r) \, d\bar{x}dxdy                         \\
&=&\sum_{\mu\nu}D_{\sigma\mu\nu}\frac{e^{-4c_{\nu}(\rho -
\rho_a)}e^{-4c_{\mu}\rho_a}}{\l(\mbox{dim }\nu\r)^{2g - 2}}\,
\frac{\l(\mbox{dim } \mu\r)}{\l(\mbox{dim } \nu\r)}\, \frac{\chi_{\nu}(u
^{-1})}{\l(\mbox{dim }\nu\r)}, 
\eeas
where $\rho_a$ is the area enclosed by the contractible loop.

Given the ability to disentangle the topological sectors, it is an
amusing exercise to compute the expected value of the topology of a
random bundle. That is, let $f: \Gamma \rightarrow R$, and, viewing
$f$ as a
map from topological sectors to the reals, define its Yang-Mills
expectation taken over bundles of all topological types ($G$ and $g$
are fixed) by
\be                                             \label{topexp}
\l\< f \r\> = \frac{\displaystyle \sum_{u \in \Gamma}
f(u)Z(u)}{\displaystyle \sum_{u \in \Gamma} Z(u)}. 
\ee

To evaluate this expression, re-write Eq.~\ref{partitionfcn} as
\[
Z(u) = \sum_{\mu} \frac{e^{-4c_{\mu}\rho}}{\l(\mbox{dim }\mu\r)^{2g -
2}}\, \lambda_{\mu}(u ^{-1}),
\]
where $\lambda_{\mu}(u) \equiv \frac{\chi_{\mu}(u)}{\mbox{dim }\mu}$.
Note that $\lambda_{\mu}$ is a character for the representation $\mu$
of $\Gamma$. Let $\mbox{Rep }\widetilde{G}$ denote the set of
equivalence classes of irreducible representations of $\widetilde{G}$,
and, for fixed $\alpha \in \mbox{Rep }\widetilde{G}$, let $(\mbox{Rep
  }\widetilde{G})_{\alpha}$ denote the set of representations which
agree with $\alpha$ on $\Gamma$. Expanding $f: \Gamma \rightarrow R$
in characters as $f(u) = \sum_{\alpha \in \mbox{\scriptsize Rep
    }\widetilde{G}} f_\alpha \lambda_{\alpha}(u)$, and letting $\alpha
  = 0$ denote the trivial representation of $\widetilde{G}$, we shall
  prove
\begin{corollary}                       \label{topexpcor}
\[
\l\< f\r\> = \sum_{\alpha \in \mbox{\scriptsize Rep
    }\widetilde{G}}f_\alpha \,\frac{\displaystyle \sum_{\mu \in
(\mbox{\scriptsize Rep }\widetilde{G})_{\alpha}}\frac{e^{-4c_{\mu}\rho}}
{\l(\mbox{dim }\mu\r)^{2g - 2}}}{\displaystyle \sum_{\mu \in
(\mbox{\scriptsize Rep }\widetilde{G})_{0}}\frac{e^{-4c_{\mu}\rho}}
{\l(\mbox{dim }\mu\r)^{2g - 2}}}.
\]
\end{corollary}
\begin{proof}
The character $\lambda_{\mu}$ satisfies the orthogonality relation
\[
\sum_{u \in \Gamma}\lambda_{\mu}(u)\lambda_{\nu}(u ^{-1})
=\l\{\begin{array}{rl}
\#\Gamma & \mbox{if $\lambda_{\mu}(u) = \lambda_{\nu}(u)$ for
all $u \in \Gamma$}                                             \\
0 & \mbox{otherwise}    
\end{array}\r. .
\]
Thus the $\alpha$-th component of the numerator in Eq.~\ref{topexp} is
\[
\sum_{u \in \Gamma}\lambda_{\alpha}(u)Z(u) = \sum_{\mu \in
\mbox{\scriptsize Rep
}\widetilde{G}}\frac{e^{-4c_{\mu}\rho}}{\l(\mbox{dim }\mu\r)^{2g -
2}}\sum_{u \in \Gamma}\lambda_{\alpha}(u)\lambda_{\mu}(u ^{-1}) =\,
\# \Gamma\sum_{\mu \in (\mbox{\scriptsize Rep
}\widetilde{G})_{\alpha}}\frac{e^{-4c_{\mu}\rho}}{\l(\mbox{dim
}\mu\r)^{2g - 2}}.
\]
Moreover, $1 = \lambda_{0}(u)$, so the denominator in Eq.~\ref{topexp} is
the same expression with $\alpha$ replaced by $0$.
\end{proof}

\begin{remark} Since $(\mbox{Rep }\widetilde{G})_{0} = \mbox{ Rep
}G$, the evaluation of the denominator proves the statement that naively
applying the results of Refs.~\cite{me1} and \cite{me2} to topologically
non-trivial bundles is equivalent to summing over topologies.
\end{remark}

\begin{remark} Let $G=SO(3)=SU(2)/\Gamma$ for $\Gamma = \{{\bf 1}, {\bf
-1}\}$, and $f(\pm {\bf 1}) = \pm 1$. Labelling the irreducible
representations of $SU(2)$ by their dimensions, which span the positive
integers, the odd-integer representations of $SU(2)$ are trivial on
$\Gamma$, while the even-integer representations are not (that is,
$\lambda_{n}(-{\bf 1}) = (-1)^{n+1}$). With the conventions of
Ref.~\cite{me1}, the Casimir is $c(\mu) =
\frac{1}{8}(n^2 -1)$. Thus, the expected value of the topology of an
$SO(3)$-bundle is
\[
\l\< f \r\> = \frac{\displaystyle \sum_{n
\mbox{\scriptsize\hspace{.5em}even}} 
\frac{e^{-\frac{1}{2} (n^2 - 1)\rho}}{n^{2g - 2}}}{\displaystyle
\sum_{n \mbox{\scriptsize\hspace{.5em}odd}}
\frac{e^{-\frac{1}{2} (n^2 - 1)\rho}}{n^{2g - 2}}}.
\]
\end{remark}
\renewcommand{\thesection}{\Roman{section}}
\section{Measures on the moduli space of flat connections}
\label{measures}
\renewcommand{\thesection}{\arabic{section}}
Let ${\cal M}_{m}$ denote the space of flat connections modulo gauge
transformations. As Witten describes in Ref.~\cite{witten1}, there is a
natural symplectic form $\omega$ on ${\cal M}_{m}$ which defines a measure
$\mu_\omega = \frac{1}{\#Z(G)}\frac{\omega^n}{n!}$. Here $n =
\frac{1}{2} \dim {\cal M}_{m}$. Sengupta has shown in Ref.~\cite{sengupta}
that the small-volume limit of the Yang-Mills measure $\mu$ on ${\cal A /
G}_{m}$ described above defines, at least in genus $0$, a second measure
$\mu_0$ on ${\cal M}_{m}$.  Witten shows these two measures agree on the
total volume of ${\cal M}_{m}$, and in Ref.~\cite{forman} Forman shows that,
in fact, $\mu_\omega = \mu_0$.

The picture of ${\cal A / G}_{m}$ as a bundle over
$Path^{2g}\,\widetilde{G}$ suggests a new measure on ${\cal M}_{m}$. First
note that two points of ${\cal M}_{m}$ cannot lie in the same fiber, as
there is no $1$-form $\tau \in \kerpr$ for which $D_{\! A}\tau = 0$.  Thus,
the restriction of $\xi$ to ${\cal M}_{m}$, henceforth denoted
$\l.\xi\r|_{{\cal M}_{m}}$, is invertible. Moreover, $\xi({\cal M}_{m}) =
\l\{\vec{\gamma} \in Path^{2g}\,\widetilde{G} : \gamma_{i} \mbox{ is
constant }\r\}$. (This is another way of saying that ${\cal M}_{m}$ may be
viewed as the representations of $\pi_1(M)$ on $\widetilde{G}$.) In short,
$\l.\xi\r|_{{\cal M}_{m}}$ provides an isomorphism between ${\cal M}_{m}$
and $\widetilde{G}^{2g}_{z}$, the space of constant $2g$-tuples in
$Path^{2g}\,\widetilde{G}$. In Theorem 4.2 of Ref.~\cite{me2}, we exhibited
a global section $\sigma : Path^{2g}\,G
\rightarrow {\cal A / G}_{m}$. Here we note the restriction of
$\sigma$ to $\widetilde{G}^{2g}_{z}$ is
$\l.\xi\r|_{{\cal M}_{m}} ^{-1}$. This is an immediate consequence of the
fact that, in a given fiber, the connection $A$ representing a point
in the image of $\sigma$ is determined up to gauge transformation by
the condition that $\l< F_{A},D_{\! A}
\tau\r>$ vanishes for all $\tau \in \kerpr$. 

To define a measure on ${\cal M}_{m}$, use the Haar measure on $G$ to
define a measure on $\widetilde{G}^{2g}_{z}$ and then use $\sigma$ to
push this measure foward to ${\cal M}_{m}$. In detail, the Haar measure on $G$
defines a measure $\mu_{H}$ on $\widetilde{G}^{2g}_{z}$ which is the
product of Haar measures on the components of $\widetilde{G} ^{2g}$,
conditioned to satisfy the endpoint relation.
That is,
\[
 \mu_{H}(x_1,y_1, \cdots, x_g, y_g) = \delta \l(\prod_{i=1}^{g} x_i y_i
x_i ^{-1} y_i ^{-1}z ^{-1}\r) \, dxdy,
\]
where $\delta$ denotes the Dirac delta distribution massed at the identity
of $\widetilde{G}$.

Under the isomorphism $\l.\xi\r|_{\cal
M}$, $\mu_{H}$ defines a measure $\mu_{\xi}$ on ${\cal M}_{m}$.
As a
measure on functions on ${\cal M}_{m}$, $\mu_{\xi}$ is given as:
\be                                     \label{def}
\int_{{\cal M}_{m}} f \mu_{\xi} = \int_{\widetilde{G}^{2g}} f \comp
\sigma(\vec{x}, 
\vec{y}) 
\, \mu_{H}.
\ee
Note that the right-hand side is an obvious measure to define on ${\cal
M}_{m}$ viewed as representations of $\pi_1(M)$. This measure may be 
normalized to define $\l< f \r>_{\mu_{\xi}}$, the expectation of a function
on ${\cal M}_{m}$.

In Ref.~\cite{ks}, King and Sengupta arrive at the measure $\mu_{\xi}$ by a
construction analogous to the construction of $\xi$ reviewed in
Sec.~\ref{review}. (They, however, restrict from the outset to
connections representing elements of ${\cal M}_{m}$.) They then argue
directly that $\mu_\xi = \mu_\omega$. Here, by contrast, we will
 show that $\mu_\xi = \mu_0$. More precisely,
\begin{theorem}                         \label{onemeasure}
Up to normalization, $\mu_{\xi}$ = $\mu_{0}$ on functions which are
analytic along the fibers of ${\cal A / G}_{m}$.
\end{theorem}
\begin{proof}
We shall show  
$\lim_{\rho \rightarrow 0} \l< f \r>_{\mu} = \l< \l.f\r|_{{\cal M}_{m}}
\r>_{\mu_{\xi}}$. 
Writing out the path integral expression for $\l< f \r>_{\mu}$ and
performing the Gaussian integral over the fibers yields
\[
\l< f \r>_{\mu} =
\frac{1}{Z}\int_{Path^{2g}\,\widetilde{G}}\hat{f}(\vec{\gamma})\,
\xi_{*}(\mu) (\vec{\gamma})
\]
where $\hat{f}$ is obtained from $f\l(\sigma(\vec{\gamma}) + \tau\r)$ by
performing the Gaussian integral over $\tau \in \kerpr$. Since $\xi_{*}(\mu)
$ is the product of Wiener measures, and the latter are determined by their
behavior on cylinder sets, we may assume, without loss of generality, that
$\hat{f}(\vec{\gamma})$ depends on $\alpha_i$ and $\beta_i$ evaluated at
points $p_{ij}$ and $q_{ik}$, respectively, where $j = 1,2, \cdots, m_i$ and
$k = 1,2, \cdots, n_i$ (and $i= 1,2, \cdots, g$). Then, as in the examples
of Sec.~\ref{review},
\bea
\lefteqn{\l< f \r>_{\mu} = \l< \hat{f} \l(\alpha_1(p_{11}),
\alpha_1(p_{12}),\cdots, 
\beta_{g}(q_{gn_{g}})\r)\r>_{\xi_{*}(\mu)}=}       \nonumber	\\ 
& &\frac{1}{Z}\int \hat{f}(x_{11}, x_{12}, \cdots, y_{gn_{g}}) 
\prod_{j=1}^{m_{i}+1} H\l( x_{i(j-1)}^{-1} x_{ij}; \Delta t_{ij}\r)\nonumber\\
& &\times \prod_{k=1}^{n_{i}+1} H\l( y_{i(k-1)}^{-1} y_{ik}; \Delta
s_{ik}\r) \delta\l(\prod_{i=1}^{g}x_{i0}y_{i(n_{i}+1)}^{-1}
x_{i(m_{i}+1)}^{-1}   y_{i0} z ^{-1}\r) \, \prod_{i,j,k} 
dx_{ij}dy_{ik},						\label{fexp} 
\eea
where $\Delta t_{i0} = 0$, $\Delta t_{ij}$ is twice the area between the
radial loops through $p_{ij}$ and $p_{i(j-1)}$ and $\Delta s_{ik}$ is defined
similarly. As $\rho$ approaches $0$, so do $\Delta t_{ij}$ and $\Delta
s_{ik}$. However, as $\Delta t$ approaches $0$,
$H(x;\Delta t)$ becomes a delta function massed at $x$. Thus, if we may take
the limit prior to integrating, 
\beas
\lefteqn{\lim_{\rho \rightarrow 0}  \l< f \r>_{\mu} =}     	\\
& & \frac{1}{Z}\int \hat{f}(x_{11}, x_{12}, \cdots, y_{gn_{g}}) 
\prod_{j=1}^{m_{i}+1} \delta\l( x_{i(j-1)}^{-1} x_{ij}\r)       \\
& & \times \prod_{k=1}^{n_{i}+1} \delta\l( y_{i(k-1)}^{-1} y_{ik}\r) 
\delta\l(\prod_{i=1}^{g}x_{i0}y_{i(n_{i}+1)}^{-1} x_{i(m_{i}+1)}^{-1}
y_{i0} z ^{-1}\r)
\, \prod_{i,j,k} dx_{ij}dy_{ik}.
\eeas
Performing the integrations over $x_{ij}$ and $y_{ik}$ for $j,k \neq
0$ changes all the $x_{ij}$'s and $y_{ik}$'s to $x_{i0}$'s and
$y_{i0}$'s, respectively, leaving
\bea
\lefteqn{\lim_{\rho \rightarrow 0}  \l< f \r>_{\mu} =}  \nonumber   \\
& & \frac{1}{Z}\int_{\widetilde{G}^{2g}}
\hat{f}(x_{10}, x_{10}, \cdots, y_{g0}) 
\delta\l(\prod_{i=1}^{g}x_{i0}y_{i0}^{-1} x_{i0}^{-1}
y_{i0} z ^{-1}\r)
\, \prod_{i} dx_{i0}dy_{i0}.                            \label{fresult}
\eea
The right-hand side is exactly $\frac{1}{Z}\int
\l.\hat{f}\r|_{\xi({\cal M}_{m})} \, \mu_{H}$. The assumption we made
about interchanging limit and integration is
\[
\lim_{t \rightarrow 0} \int f(x) H\l(y ^{-1} x; t\r) \, dx = \int f(x)
\delta\l(y ^{-1} x \r) \, dx.
\]
That each side is equal to $f(y)$ follows from the definition
of the heat kernel (and the continuity in $t$ of solutions of the heat
equation) on the left and the definition of the distribution $\delta$
on the right. We thus have 
\[
\lim_{\rho \rightarrow 0}  \l< f \r>_{\mu} = \frac{1}{Z}\int
\l.\hat{f}\r|_{\xi({\cal M}_{m})} \, \mu_{H}.
\]  
According to Eq.~\ref{def}, we must now  show
\be                                     \label{fisnice}
\lim_{\rho \rightarrow 0}\hat{f} = \l.f\r|_{{\cal M}_{m}} \comp \sigma.
\ee
First, note that if $f$ is constant along the fibers, that is, if
$f\l(\sigma(\vec{\gamma}) + \tau\r)$ = $f \comp \sigma$, then, for
any $\rho$, $\hat{f} = f \comp \sigma$. More generally,
if $f\l(\sigma(\vec{\gamma}) + \tau\r)$ is an $n$th order
polynomial in $\tau$, then $\hat{f}$ is an $n$th order polynomial in
$\rho$, whose constant term is $f \comp \sigma$. This follows from
standard manipulations of Gaussian integrals and the
fact that, in two dimensions, the area $\rho$ plays the role of the
coupling constant. Thus, for $f$ 
polynomial, or, more generally, analytic, in $\tau$,  Eq.~\ref{fisnice}
holds (with the convergence being uniform); hence, up to normalization,
$\mu_{\xi}=\mu_{0}$ on analytic functions. 
\end{proof}
\begin{remark}The restriction to analytic functions of the fiber is not
terribly severe. In most field theory, polynomials are sufficient.
Moreover, the freedom
in choosing the fundamental domain is sufficient to ensure that a
large class of Wilson lines may be realized as functions which
are {\em constant} on each fiber.
\end{remark}
Theorem~\ref{onemeasure} provides a new proof of Forman's
generalization of Sengupta's result:
\begin{corollary}
The small-volume limit of $\mu$ is supported on ${\cal M}_{m}$.
\end{corollary}

\noindent Observe that only the restriction of
$f$ to ${\cal M}_{m}$ enters into the above calculation of $\l< f
\r>_{\mu_{0}}$. Specifically, let $\chi_{R}$ be the indicator function of a
measurable set $R \subset {\cal A / G}_{m}$ for which $R \cap {\cal M}_{m} =
\emptyset$, and let $\chi_{R}^{smooth}$ be a smooth, non-negative function
which is $1$ on $R$ and has support in the complement of ${\cal M}_{m}$.
Now, let $\l\{\chi^{n}_{R}\r\}$ be a sequence of analytic functions
converging uniformly to $\chi_{R}^{smooth}$. Then, by the theorem, $\l<
\chi^{n}_{R}\r>_{\mu_{0}} = \l< \l.\chi^{n}_{R}\r|_{{\cal
M}_{m}}\r>_{\mu_{\xi}}$.  By the construction of the sequence, there is some
$N$ such that $\chi^{n}_R$ vanishes on ${\cal M}_{m}$ for all $n > N$.
Hence, $\l<\chi^{smooth}_R \r>_{\mu_{0}} = 0$ and thus $\l<\chi_R
\r>_{\mu_{0}} = 0$.  

\noindent\begin{remark} We have assumed the existence of the sequence
$\l\{\chi^{n}_{R}\r\}$. If ${\cal A / G}_{m}$ were a
finite-dimensional manifold, the Stone-Weierstrass Theorem would
ensure the existence of such a sequence, but at present the above ``proof'' of
the corollary is a heuristic argument. 
\end{remark} 

Comparing the arguments to $\hat{f}$ in Eq.~\ref{fexp} and
Eq.~\ref{fresult} shows that one effect of going to the small-volume
limit is to project from $Path^{2g}\,\widetilde{G}$ to
$\widetilde{G}^{2g}_{z}$ by evaluating each component path at $t=0$.
Denote this evaluation map by $e: Path^{2g}\,\widetilde{G} \rightarrow
\widetilde{G}^{2g}_{z}$. This projection suggests another measure
$\mu_{e}$ on ${\cal M}_{m}$ which is the push-forward of $\mu$ by the
projection from ${\cal A / G}_{m}$ to ${\cal M}_{m}$ given by $\sigma \comp
e \comp \xi$. That is, 
\[
\mu_{e} = \sigma_* e_* \xi_*(\mu).
\]
As $\sigma$ is an isomorphism and $\xi_*(\mu)$ is the product of
Wiener measures described in Sec.~\ref{top}, the only new
feature is the effect of pushing forward by $e$. This means
integrating using the Wiener measures with fixed left end-points.
The derivation of Eq.~\ref{fixedleft}, with minor modifications to work
in a fixed topological sector and to integrate only over right
end-points, leads to 
\[
\l< f \r>_{\mu_{e}} = \frac{1}{Z} \int f \comp \sigma (x_{1}, \cdots
y_{g}) H\l(\prod_{i=1}^{g}x_{i}y_{i}x_{i}^{-1}y_{i}^{-1};2\rho\r)
\, dx_{i}dy_{i}.
\]
Comparing this with $\l< f \r>_{\mu_{\xi}}$ ({\em c.f.} Eq.~\ref{def}),
it is clear that, up to normalization,
\[
\mu_{e} =
\frac{H\l(\prod_{i=1}^{g}x_{i}y_{i}x_{i}^{-1}y_{i}^{-1};2\rho\r)}
{\delta\l(\prod_{i=1}^{g}x_{i}y_{i}x_{i}^{-1}y_{i}^{-1};2\rho\r)} 
\mu_{\xi}.
\]
The measure $\mu_{e}$ is thus a new measure on ${\cal M}_{m}$ which agrees
with $\mu_{\xi}$ only in the limit as $\rho$ approaches $0$.  

\renewcommand{\thesection}{\Roman{section}}
\section{Conclusion}
\renewcommand{\thesection}{\arabic{section}}

We have extended the path integral formulation of Yang-Mills on
Riemann surfaces to treat each topological sector separately. The
result is in agreement with Witten's approach and is sufficiently
explicit to compute quantum expectations of a large class of Wilson
lines. It  also provides a new measure on ${\cal M}_{m}$, the moduli
space of flat connections.

There are many routes to defining measures on ${\cal M}_{m}$. The symplectic
form $\omega$ on ${\cal M}_{m}$ and the small-volume
limit of $\mu$ define the measures $\mu_{\omega}$ and $\mu_{0}$,
respectively. The view of ${\cal 
A / G}_{m}$ as a bundle over $Path^{2g}\,G$ suggests the measures
$\mu_{\xi}$ and $\mu_{e}$.
However, this apparent profusion of measures on ${\cal M}_{m}$ is
in fact a pair of measures. Combining Theorem~\ref{onemeasure} with 
Forman's result:
\[
\mu_{\xi} = \mu_{0} = \mu_{\omega}.
\]
\section*{Acknowledgements}
The author wishes to thank K. Morrison for asking (perhaps
rhetorically) the question answered in Corollary~\ref{topexpcor}, and
to thank the referee for bringing to the author's attention the papers cited
in Refs.~\cite{senguptapreprint} and~\cite{senguptanato}.

This material is based upon work supported in part by the National
Science Foundation under Grant \#DMS-9307608.

\end{document}